\documentclass[useAMS,usenatbib]{mn2e}

\usepackage{graphicx}
\usepackage{amsmath}
\usepackage{amssymb}

\title[Anisotropic CR diffusion]{Anisotropic CR diffusion and $\gamma$-ray production close to supernova remnants, with an application to W28}
\author[L. Nava and S. Gabici]
{L. Nava$^{1}$\thanks{E-mail: lara.nava@apc.univ-paris7.fr} 
and S. Gabici$^{1}$\\
$^{1}$APC, AstroParticule et Cosmologie, Universit\'e Paris Diderot, CNRS/IN2P3, CEA/Irfu, Observatoire de Paris, Sorbonne Paris Cit\'e,\\ 10, rue Alice Domon et L\'eonie Duquet, 75205 Paris Cedex 13, France}
\begin{document}

\date{}

\pagerange{\pageref{firstpage}--\pageref{lastpage}} \pubyear{}

\maketitle

\label{firstpage}

\begin{abstract}
Cosmic rays that escape their acceleration site interact with the ambient medium and produce gamma rays as the result of inelastic proton--proton collisions.
The detection of such diffuse emission may reveal the presence of an accelerator of cosmic rays, and also constrain the cosmic ray diffusion coefficient in its vicinity.
Preliminary results in this direction have been obtained in the last years from studies of the gamma--ray emission from molecular clouds located in the vicinity of supernova remnants, which are the prime candidate for cosmic ray production. Hints have been found for a significant suppression of the diffusion coefficient with respect to the average one in the Galaxy. However, most of these studies rely on the assumption of isotropic diffusion, which may not be very well justified. 
Here, we extend this study to the case in which cosmic rays that escape an accelerator diffuse preferentially along the magnetic field lines. As a first approximation, we further assume that particles are strongly magnetized and that their transport perpendicular to the magnetic field is mainly due to the wandering of the field lines. The resulting spatial distribution of runaway cosmic rays around the accelerator is, in this case, strongly anisotropic.
An application of the model to the case of the supernova remnant W28 demonstrates how the estimates of the diffusion coefficient from gamma--ray observations strongly depend on the assumptions made on the isotropy (or anisotropy) of diffusion. For higher levels of anisotropy of the diffusion, larger values of the diffusion coefficient are found to provide a good fit to data. Thus, detailed models for the propagation of cosmic rays are needed in order to interpret in a correct way the gamma--ray observations.
\end{abstract}


\section{Introduction}

Galactic Cosmic Rays (CRs) are mainly constituted by relativistic protons and are believed to be accelerated at SuperNova Remnant (SNR) shocks via first order Fermi mechanism \citep{hillas}. Though very popular, this scenario still needs to be conclusively proven by observations. 

If CRs are indeed accelerated at SNRs, these objects must be gamma--ray sources. This is because the CRs accelerated at the shock undergo inelastic proton--proton interactions with the ambient medium and produce neutral pions which in turn decay into gamma rays \citep{dav,nt}. Several SNRs have been detected in gamma rays at both TeV \citep[e.g.][]{jim} and GeV \citep[e.g.][]{giordano} energies, in agreement with such expectations. However, it is often difficult to determine whether the origin of the gamma--ray emission is hadronic, and thus related to the acceleration of CRs, or due to leptonic mechanisms such as inverse Compton scattering. 
For this reason, multi-wavelength studies of SNRs have been extensively carried out in an attempt to solve this degeneracy.
Though for some individual SNRs it has been possible to ascribe the gamma--ray emission exclusively and quite confidently to hadronic \citep[e.g.][]{fermitycho,giovanni} or leptonic \cite[e.g.][]{fermiRXJ,donRXJ} processes, in other cases this ambiguity remains a problem.

An alternative way to reveal the presence of a CR source is by searching for the radiation produced by CRs that escape the acceleration site \citep{atoyan, gabici07,rodriguez,gabici09}. At some stage of the dynamical evolution of the SNR, CRs are expected to leave the shock region and escape into the interstellar medium. The details of the escape mechanism are still not very well understood \citep[see e.g.][and references therein]{gabiciescape}, but it is generally believed that the ability of a SNR in confining particles decreases gradually with the shock speed, with higher energy particles leaving the shock earlier than low energy ones. Once escaped, CRs diffuse away from the SNR and produce gamma--rays in interactions with the ambient gas. To date, some possible evidence for particle escape from SNRs has been pushed forward by the observation of diffuse gamma--ray emission from the vicinity of the shell of the SNRs W28 \citep{W28hess,W28agile,W28fermi} and W44 \citep{W44fermi}. In both cases, the emission is clearly located outside of the shell and it is spatially coincident with the location of massive Molecular Clouds (MCs). This would favor a scenario in which the MCs are illuminated  by the runaway CRs and, being very massive, become prominent gamma--ray sources \citep[for the case of W28 see e.g.][and discussion in Section~\ref{sect:W28}]{gabiciW28}.

Besides revealing the presence of a CR source, the gamma--ray emission from runaway CRs can also be used to constrain the particles' diffusion coefficient in the region surrounding the accelerator. This is very important for several reasons: first of all, a theoretical determination of the diffusion coefficient is a very complex task \citep[see e.g.][]{lazarian,lazarian08} and observational constraints are needed in order to guide and constrain models. In addition to that, the diffusion of CRs is believed to be a non--linear process in which CRs themselves generate via streaming instability the turbulence they scatter off \citep[e.g.][]{kulsrud}. This is particularly relevant close to CR sources, where the CR density is expected to be very high, and possibly sufficient to suppress significantly the diffusion coefficient through streaming instability \citep{plesser,malkov}. Thus, an empirical determination of the diffusion coefficient can reveal precious information on the ways in which particles and waves interact in astrophysical plasmas.

Most of the studies aimed at the determination of the CR diffusion coefficient from gamma--ray observations rely on the assumption of {\it isotropic diffusion} \citep{diego,fujita,li,gabiciW28,ohira,lazarianW28}. The common rationale of these approaches can be summarized as follows: if SNRs are the sources of CRs, they have to convert a fraction $\eta \approx 10\%$ of their explosion energy $E_{SN} = 10^{51} E_{51} {\rm erg}$ into accelerated particles. If the diffusion of CRs proceeds isotropically, after a time $t$ from escape CRs of a given energy $E$ are distributed roughly homogeneously within a distance $R_d(E) \approx \sqrt{6\,D(E) \times t}$ from the SNR. Here, $D(E)$ is the energy dependent diffusion coefficient of CRs. Gamma--ray observations of MCs located in the vicinity of W28 or W44 tells us which is the CR density $n_{CR}(E)$ needed to explain the observed emission. According to what said above, such density has to be of the order of $n_{CR}(E) \approx f_{sp}(E) ~ \eta E_{SN}/R_d^3$, where the factor $f_{sp}(E)$ contains the information on the shape of the spectral energy distribution of escaping CRs. For aged SNRs such as W28 and W44, the time $t$ after the escape can be identified with the SNR age $t_{age}$, and thus an expression for the diffusion coefficient can be obtained, and reads: $D \approx (f_{sp} \eta E_{SN}/n_{CR})^{2/3} t_{age}^{-1}$. Since the values of all the physical quantities present on the right side of the equation can be inferred from observations, the expression provides a direct estimate of the diffusion coefficient.

As an example, we summarize here the results obtained by \citet{gabiciW28} in interpreting the gamma--ray emission observed from the MCs located close to the SNR W28. They obtained a good fit to the gamma--ray spetrum measured by H.E.S.S. at photon energies $\gtrsim 300$~GeV by assuming a diffusion coefficient for $\approx 3$~TeV CRs of the order of $D(3~{\rm TeV}) \approx 5 \times 10^{27} (\eta/0.1)^{2/3}$~cm$^2$/s. The corresponding diffusion length $R_d$ of these particles is of the order of 100 pc. Also the broad band gamma--ray spectrum from GeV to TeV energies can be fitted by adjusting the energy dependence of the diffusion coefficient and the distances between the SNR and the clouds. The important point here is the fact that the estimated diffusion coefficient is more than one order of magnitude smaller than the one normally adopted to describe the propagation of CRs of energy $\gtrsim$~TeV in the galactic disk, which is $\approx 10^{29}\,$cm$^2$/s \citep{andyreview,fiorenzareview}.
These results are very similar to the ones obtained by other authors by means of similar modeling \citep{fujita,li,ohira,lazarianW28} and seem to point toward a drop of the diffusion close to the SNR W28.

However, the validity of the assumption of isotropic diffusion of CRs needs to be discussed. In fact, if the intensity of the turbulent field $\delta B$ on scales resonant with the Larmor radius of particles is significantly smaller than the mean large scale field $B_0$ (i.e. if $\delta B/B_0 \ll 1$), then {\it CR diffusion becomes anisotropic}, with particles diffusing preferentially along the magnetic field lines \citep[e.g.][]{fabien}. In the limiting (but still reasonable) case in which the perpendicular diffusion coefficient can be set equal to zero, the transport of CRs across the mean field is mainly due to the wandering of magnetic field lines \citep{parker}. This is the situation that we investigate in this paper.

To give a qualitative idea of the role that anisotropic diffusion can play in these study, let us consider an idealized case in which particles that escape a SNR diffuse along a magnetic flux tube characterized by a very long coherence length (i.e. the magnetic flux tube is preserved for a long distance). In this case, after a time $t$ particle will diffuse up to a distance $R_d \approx \sqrt{2\,D_\parallel \times t}$ along the tube (here $D_\parallel$ is the {\it parallel} diffusion coefficient), while their transverse distribution will be equal to the radius of the SNR shock at the time of their escape, $R_{sh}$, which is of the order of $\approx$~1--10 pc. Thus, the enhanced CR density in the flux tube will be proportional to $n_{CR} \propto (R_d R_{sh}^2)^{-1}$ instead of $\propto R_d^{-3}$ as in the isotropic case. It is easy to see that the estimates of the diffusion coefficient based on the two opposite assumptions of isotropic and one-dimensional diffusion will differ by a factor of $\approx (R_d/R_{sh})^{4/3}$, which can be much larger than an order of magnitude! Thus, it is of paramount importance to investigate how the interpretation of gamma--ray observations depends on the assumptions made concerning CR diffusion.

In Section~\ref{sect:model} we develop a model for CR propagation in which CRs are strongly magnetized and diffuse uniquely along the magnetic field lines. The wandering of the field lines is also taken into account, and a diffusion coefficient $D_m$ for the magnetic field lines that depends on the properties of the turbulent field is defined \citep[see e.g.][]{duffy}. In Section~\ref{sect:results} the model is used to predict the spatial distribution of runaway CRs and their spectrum. Finally, we apply the model to the case of the SNR W28 in Section~\ref{sect:W28}. A good fit to data is obtained, and the estimate of the parallel diffusion coefficient is found to depend on the level of anisotropy of the diffusion. For higher level of anisotropy, i.e. smaller values of $D_m$, larger values of the particles' diffusion coefficient are needed in order to fit data. A discussion of the results and of future perspectives in this line of research can be found in Section~\ref{sect:conclusions}.

\section{Cosmic--ray transport in the presence of magnetic field line wandering}
\label{sect:model}

Consider a magnetic flux tube whose mean magnetic field $B_0$ is assumed to lie along the $z$-axis, perpendicular to the $(x,y)$ plane. The wandering of magnetic field lines is due to long wavelength perturbations, i.e. perturbations on scales much larger than the particles' gyroradii, with root mean square amplitude $\delta B$.
The condition for the validity of quasi--linear theory is $\delta B/B_0 \ll L_{\perp}/L_{\parallel}$, where $L_{\perp}$ and $L_{\parallel}$ are the field coherence lengths perpendicular and parallel to $B_0$, respectively \citep{kadomtsev}. 
According to quasi--linear theory, the field lines passing in the vicinity of $(x_0,y_0)$ at $z = 0$ are spread over a larger region as they reach a given $z$.
The probability distribution describing this spreading of field lines is gaussian and characterized by $\langle (x-x_0)^2 \rangle = \langle (y-y_0)^2 \rangle = 2 D_m z$, where brackets indicate an ensemble average and $D_m$ is a diffusion coefficient for field lines \citep{parker}.
For a broad band Fourier spectrum of the perturbation, the coherence lengths can be expressed as $L_{\perp,\parallel} = 2 \pi/\Delta k_{\perp,\parallel} \approx 2 \pi / k_{\perp,\parallel}$, with $k_{\perp}$ and $k_{\parallel}$ the characteristic wave--vectors of the perturbation \citep{achterberg}. Under these circumstances, the diffusion coefficient is $D_m = (\delta B/B_0)^2 L_{\parallel}/4$ \citep{kadomtsev,duffy}. It is also possible to define the Lyapunov length $\lambda_L = L_{\perp}^2 (\delta B/B_0)^{-2}/L_{\parallel}$ which describes the exponential separation of field lines whose initial separation is smaller than $L_{\perp}$ \citep{isichenko}. This can be interpreted as the length above which the flux tube is disrupted by field line divergence. For fiducial values of the parameters  the length of the flux tube is of the order of a few hundred parsecs  \citep[see e.g.][]{plesser}.

We are interested here in studying the propagation of CRs in the presence of magnetic field line wandering. Energetic particles diffuse along and across the magnetic field line as the result of resonant interactions with magnetic field perturbations. Such perturbations are characterized by length scales of the order of the particles' Larmor radii. Such scales are much smaller than the ones responsible for field line wandering. According to quasi--linear theory, the ratio between the parallel to perpendicular diffusion coefficient is $D_{\parallel}/D_{\perp}= 1 + (\lambda_{\parallel}/r_g)^2$, where $\lambda_{\parallel}$ is the particle's mean free path along the field line and $r_g$ is its gyroradius. In the interstellar medium it is believed that $\lambda_{\parallel} \gg r_g$ which implies $D_{\perp} \ll D_{\parallel}$ \citep[e.g.][]{fabien}. Thus, in the following we will neglect the diffusion of particles perpendicular to the field lines. In other words, a given particle remains attached to the same field line. Under this conditions, in a time interval $\Delta t$ a particle diffuses along a given field line a distance $\langle (\Delta z)^2 \rangle = 2 D_{\parallel} \Delta t$, but over such a distance $\Delta z$ along the $z$--axis the field line is displaced by an amount $\langle (\Delta x)^2 \rangle = 2 D_m \Delta z$. This leads to \citep{getmantsev,rosenbluth,cptuskin}:
\begin{equation}
\langle (\Delta x)^2 \rangle \propto D_m \sqrt{D_{\parallel} \Delta t}
\end{equation}
which describes a sub--diffusive transport of particles perpendicular to the mean magnetic field $B_0$.

This behavior of energetic particles due to the combination of the particle diffusion along the field lines and the random walk of the field lines themselves has been often referred to as {\it compound diffusion}, or {\it anomalous diffusion}. Models of compound diffusion have been developed and used in a great variety of contexts, to study phenomena like the heat transport in Tokamak \citep{rosenbluth,isichenko}, the propagation of energetic particles in the solar wind \citep{parker,zimbardo}, the confinement of CRs in the Galaxy \citep{getmantsev,lingenfelter,cptuskin} and their acceleration at astrophysical shocks \citep{achterberg,duffy,john}.
In this paper, we apply the formalism of compound diffusion to another context, which is the propagation of CRs in the vicinity of their sources, i.e. SNRs, after they escape the acceleration region.
We will show that in this situation an accurate modeling is needed in order to interpret in a correct way the gamma--ray observations of MCs located in the vicinity of SNRs.

In order to describe the compound diffusion of CRs we adopt the mathematical formalism developed by \citet{webb} and we define $P_{\rm FRW} (x|z)$ as the probability to find a field line displaced by an amount $\Delta x$ after a step of length $z$ along the direction of the umperturbed field $B_0$. From what said above, it follows that:
\begin{equation}\label{eq:frw}
P_{\rm FRW} (\Delta x|z)=\frac{1}{\sqrt{4\pi D_m z}}\exp\left[-\frac{(\Delta x)^2}{4D_m z}\right]
\end{equation}
which corresponds to a diffusive behavior of field lines (FLW stands for Field line Random Walk). 
A similar equation holds for the displacement $\Delta y$.
This has to be combined with the probability $P_{\rm \parallel} (z|\Delta t)$ that a particle moves a distance $z$ along the field line in a time $\Delta t=t-t_0$.
For diffusive transport of particles along the field we have:
\begin{equation}
\label{eq:parallel}
P_{\rm \parallel} (z|\Delta t)=\frac{1}{\sqrt{4\pi D_\parallel \Delta t}}\exp\left[-\frac{z^2}{4 D_\parallel\Delta t}\right]
\end{equation}
The probability for a particle to reach the position $(x,y,z)$ at the time $t$, when its position at the time $t_0$ was ($x_0,y_0,z_0=0$), is then the product of $P_{\rm FRW}$ with $P_{\rm \parallel}$:
\begin{equation}
\label{eq:prob}
P(\Delta x, \Delta y, z; \Delta t)=P_{\rm \parallel} (z|\Delta t)\, P_{\rm FRW} (\Delta x|z)\,P_{\rm FRW} (\Delta y|z)
\end{equation}

In order to model the escape of CRs from a SNRs we assume, following \citet{plesser}, that particles are injected in the flux tube in the $xy$-plane at $z = 0$, within a circular region whose radius is equal to the SNR shock radius $R_{sh}(E)$. Since CRs of different energy are expected to escape the remnant at different times, the radius of the injection region is an energy dependent quantity. Following \citet{gabici09} we assume a power law scaling to connect the particle energy of the runaway CRs with the time after the supernova explosion:
\begin{equation}
\label{eq:escapetime}
E_{esc} = E_{MAX} \left( \frac{t_{esc}}{t_{Sed}} \right)^{-\delta}
\end{equation}
where the implicit assumption has been made that the maximum energy of CRs accelerated in a SNR $E_{MAX}$ is reached at the time $t_{Sed}$ which marks the transition between the free expansion and the Sedov phases of the SNR evolution, and that CRs are gradually released in the interstellar medium from that time on. The Sedov phase is characterized by a scaling $R_{sh} \propto t^{2/5}$ which gives:
\begin{equation}
R_{sh}(E_{esc}) \propto \left( \frac{E_{esc}}{E_{MAX}} \right)^{-\frac{2}{5 \delta}}
\end{equation}
which is what is assumed in the following. Other parameterizations of the escape time of CRs can be easily implemented.

The spatial distribution of CRs can now be obtained by integrating the probability function given by Eq.~\ref{eq:prob} within the range $R_0 = \sqrt{(x_0^2+y_0^2)} \le R_{sh}(t_{esc}(E))$. To do so, it is convenient to adopt a cylindrical coordinate system and express the Field line Random Walk part of Eq.~\ref{eq:prob} as a function of the quantities $R=\sqrt{x^2+y^2}$, $R_0 = \sqrt{(x_0^2+y_0^2)}$ and $cos(\Delta\varphi)=(xx_0+yy_0)/(RR_0)$ which leads to:
\begin{eqnarray}
f_{CR}(R,z,t,E) = A ~ \frac{E^{-\Gamma}}{\pi R_{sh}^2} ~ P_{\parallel}(z|t-t_{esc}) \times  \nonumber
\\
\times \int_0^{2 \pi} d\varphi \int_0^{R_{sh}(t_{esc})} {\rm d}R_0\,R_0\,P_{FRW}(R,R_0,\Delta\varphi|z)  
\end{eqnarray}
where it has been assumed that the total spectrum of CRs released in the interstellar medium during the whole life of the SNR is a power law $A E^{-\Gamma}$ with normalization:
\[
\begin{cases}
A=\frac{\eta E_{\rm SN} (\Gamma-2)}{E_{\rm MAX}^{2-\Gamma}}\left[ \left( \frac{E_{\rm MIN}}{E_{\rm MAX}}\right)^{2-\Gamma}-1\right]^{-1} ~~~~~~~\rm for~~\Gamma\neq2 \\
A=\frac{\eta E_{\rm SN}}{\ln(E_{\rm MAX}/E_{\rm MIN})} ~~~~~~~~~~~~~~~~~~~~~~~~~~~~~\,\rm for~~\Gamma=2
\end{cases}
\]
Here, $E_{SN}$ it the supernova explosion energy, $\eta$ is the fraction of this energy converted into CRs which are released in the interstellar medium, and $E_{MIN}$ and $E_{MAX}$ represent the extension in energy of the CR spectrum.

To describe the diffusion of CRs along field lines we adopt a diffusion coefficient which is a power-law in energy: 
\begin{equation}
\label{eq:diffusion}
D_\parallel(E)=\widetilde{D}_{\parallel}~\left(\frac{E}{10\rm\,GeV}\right)^{s}
\end{equation}
where $\widetilde{D}_\parallel$ and $s$ are considered here as free parameters.

Before proceeding in computing the spatial distribution of CRs around a SNR we notice that, for $z \ll L_{\parallel}$, Eq.~\ref{eq:frw} does not provide a good description of the field line wandering. The reason is that in this regime the lateral displacement of a field line after a step $z$ along $B_0$ is of the order of $\approx (\delta B/B_0) z = bz$, since $b = \delta B/B_0$ represents the angle between the unperturbed ($B_0$) and total ($B_0+\delta B$) magnetic field \citep{isichenko}. An accurate and quantitative analysis of the behavior of a magnetic flux tube in this regime goes beyond the scope of this paper \citep[see][for a more detailed discussion]{isichenko}. However, in order to describe this regime in a qualitative way, for $z \ll L_{\parallel}$ we substitute Eq.~\ref{eq:frw} with: 
\begin{equation}
\label{eq:conical}
P(R | z) = \frac{\vartheta\left[(b z)^2 - (R-R_0)^2\right]}{\pi (b z)^2}
\end{equation}
where $\vartheta[s]$ is the Heaviside function, equal to 1 for $s>0$ and 0 for $s<0$.
Eq.~\ref{eq:conical} roughly mimics the behavior of a flux tube characterized by a opening angle $b$.
In the intermediate region $z \approx L_{\parallel}$ we use an interpolating function to bridge the behaviors described by Eqns.~\ref{eq:frw} and \ref{eq:conical}.

\section{Results}
\label{sect:results}

In this Section we compute the spatial distribution of CRs expected in the vicinity of a SNR at a given time after the explosion.
We consider a typical supernova, characterized by the following fiducial values of parameters: an explosion energy of $E_{SN} = 10^{51}\,$erg, a mass of the ejecta equal to $M_{ej} = 1.4 \, M_{\odot}$, and a density of the circumstellar medium $n_0 = 1\,$cm$^{-3}$. We further assume that a fraction $\eta = 0.1$ of the supernova explosion energy is converted into CRs, which are injected in the interstellar medium with a power law differential energy spectrum $dN/dE \propto E^{-\alpha}$
which extends from $E_{\rm MIN}=1\,$GeV to $E_{\rm MAX}=5\,$PeV (approximately the position of the knee in the CR spectrum). 
It is known from CR data that $\alpha$ should be in the range $\approx2.1-2.4$ \citep{fiorenzareview,andyreview}.
We adopt here $\alpha=2.2$ as a representative value.
As described in Sec.~\ref{sect:model}, CRs are gradually released from the SNR during the Sedov phase that goes from $t \approx 280$~yr to $t \approx 3.6 \times 10^4\,$yr \citep{cioffi}. For the parallel diffusion coefficient of CRs (Eq.~\ref{eq:diffusion}) we assume $\widetilde{D}_\parallel = 10^{28}\,$cm$^2$/s and $s = 0.5$.

\begin{figure}
\hskip -0.38cm
\includegraphics[scale=0.52]{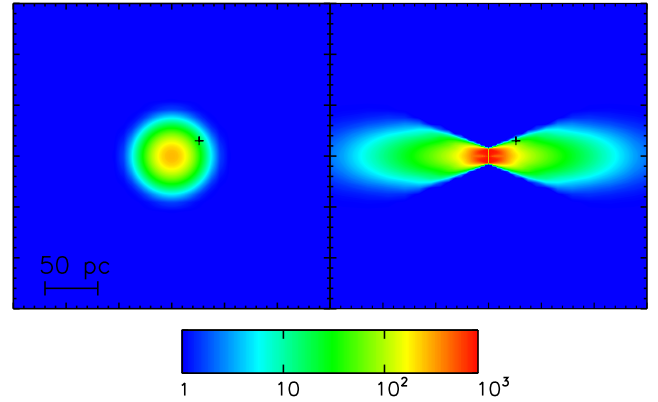}
\caption{Cosmic ray over-density around a typical supernova remnant (see text for details) for a particle energy of $E=1\,$TeV at a time $t=10\,$kyr after the explosion. The left panel refers to an isotropic diffusion coefficient of cosmic rays equal to $D = 5 \times 10^{26} (E/10~{\rm GeV})^{0.5}\,$cm$^2$/s, while the right panel refers to an anisotropic diffusion scenario with $D_\parallel = 10^{28}(E/10~{\rm GeV})^{0.5}\,$cm$^2$/s, $D_m = 1\,$pc, and $b^2 = (\delta B/B_0)^2 = 0.2$. The black cross marks the a position at which the CR over-density is equal in the two panels.}
\label{fig:iso}
\end{figure}

As a first step, we compare in Fig.~\ref{fig:iso} the results that are obtained if an isotropic diffusion coefficient is assumed \citep[as, e.g., in][]{atoyan,gabici09}, with the ones obtained for the anisotropic diffusion model that we consider in this paper. 
In both panels of Fig.~\ref{fig:iso}, the SNR is located at the centre of the field and the color code refers to the excess of CRs with respect to the average density of CRs in the Galaxy, which is \citep[e.g.][]{pdg}:
\begin{equation}
N_{CR}^{gal}(E) \approx 1.8 ~ \left( \frac{E}{\rm GeV} \right)^{-2.7} {\rm GeV^{-1}cm^{-2}s^{-1}sr^{-1}}
\end{equation}
Over-densities are plotted for a particle energy of $1\,$TeV and for a time $t = 10\,$kyr after the supernova explosion.
Here the diffusion coefficient of the magnetic field lines is set equal to $D_m = 1\,$pc, with $b^2 = (\delta B/B_0)^2 = 0.2$ (different values of $D_m$ will be explored in the following). This corresponds to a parallel coherence length of the perturbation of $L_{\parallel} = 20\,$pc.

\begin{figure*}
\centering
\includegraphics[width=1.\textwidth]{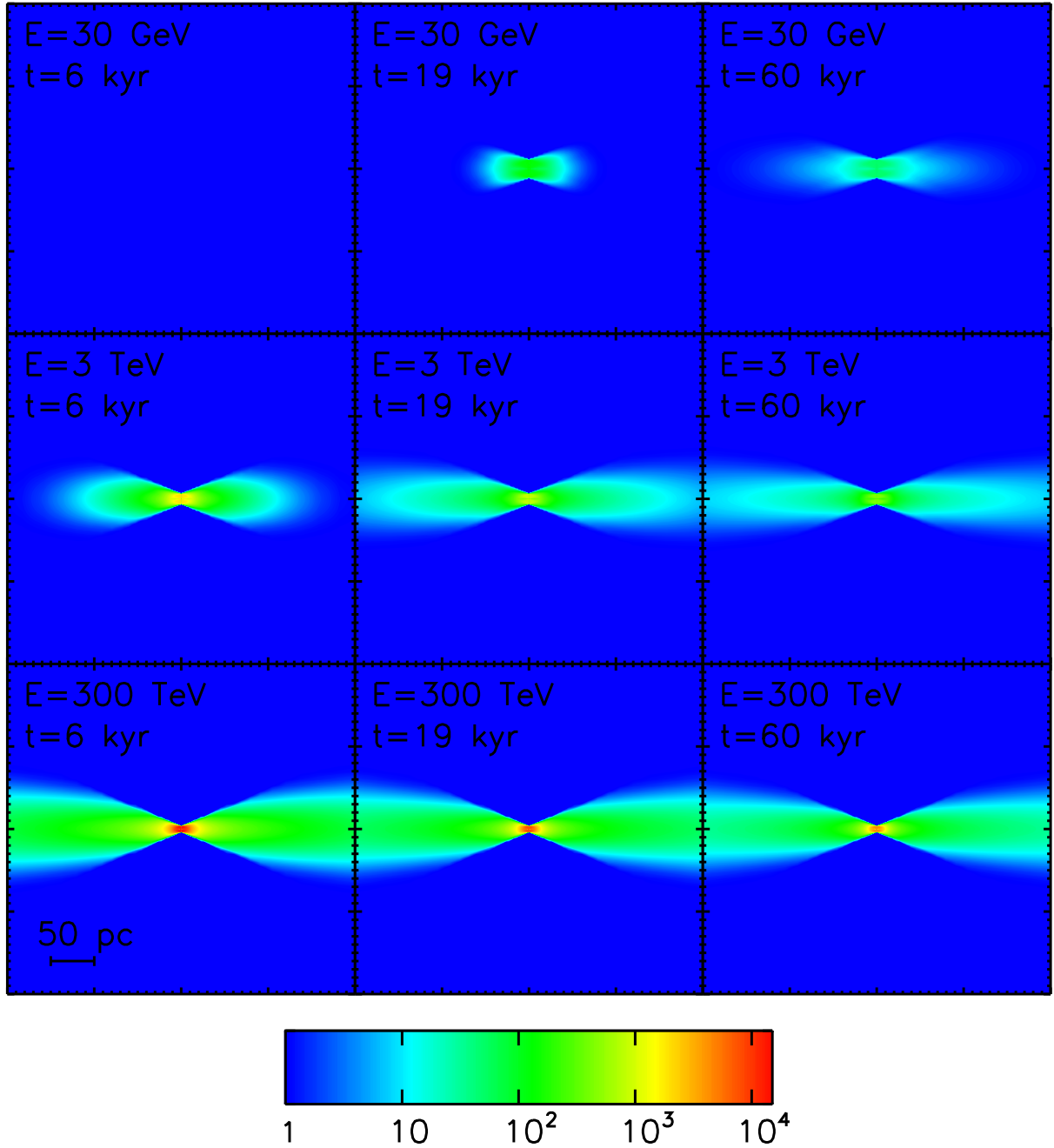}
\caption{Cosmic ray over--density with respect to the galactic background around a supernova remnant (located in the centre of the field). The particle energy is $E=30\,$GeV (upper row), $E=3\,$TeV (middle row) and $E=300\,$TeV (lower row) and the age of the supernova remnant is $t=6\,$kyr (left column), $t=19\,$kyr (middle column) and $t=60\,$kyr (right column), respectively. The diffusion coefficients are as in Fig.~\ref{fig:iso}.}
\label{fig:cr_map}
\end{figure*}

\begin{figure*}
\centering
\includegraphics[width=1.\textwidth]{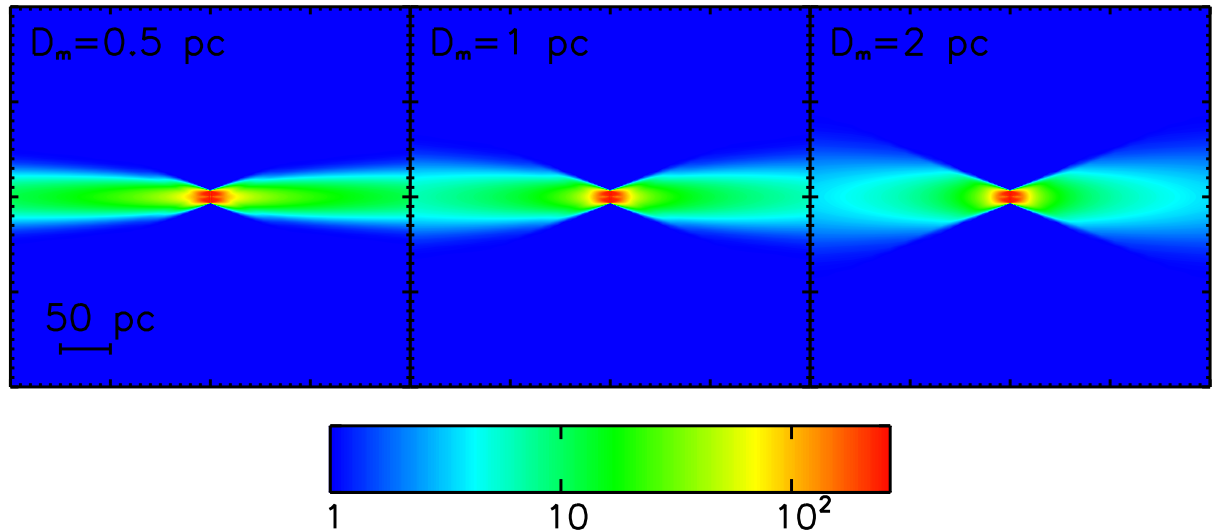}
\caption{Cosmic ray over--density for particles of 3 TeV around a supernova remnant of age 19 kyr. The assumed values of the parameters are as in Fig.~\ref{fig:cr_map}, except for the diffusion coefficient of the magnetic field lines $D_m$ which is 0.5, 1, and 2 pc (left to right panel).}
\label{fig:Dm}
\end{figure*}

The spatial distribution of CRs is strikingly different in the two scenarios: spherically symmetric in the left panel, and strongly elongated in the direction of the magnetic field flux tube in the right panel. A filamentary diffusion of CRs was also found in the numerical simulations by \citet{giacinti}.
The same parameters have been used to compute the over--densities in the two scenarios, with the exception of the CR diffusion coefficient, which in the left panel has been assumed to be isotropic and equal to $D = \widetilde{D}\,(E/10\,{\rm GeV})^{0.5}\,$cm$^2$/s with $\widetilde{D} = 5 \times 10^{26}\,$cm$^2$/s. The choice of two significantly different values for $\widetilde{D}$ and $\widetilde{D}_\parallel$, with $\widetilde{D} \ll \widetilde{D}_\parallel$ has been made in order to obtain the same level of CR over--density in the vicinity of the SNR. As an example, the black cross in Fig.~\ref{fig:iso} shows a position, located $30\,$pc away from the centre of the explosion, where the CR over-density is identical in the two panels. 
To get comparable values for the CR over--density, a much smaller (isotropic) diffusion coefficient $\widetilde{D}$ is needed in order to compensate for the larger solid angle over which CRs can propagate.
As already stressed in the introduction, this fact must be taken into account when interpreting the gamma--ray observations of molecular clouds illuminated by CRs escaping from SNRs. This will be discussed in Sec.~\ref{sect:W28}, when the model developed here will be applied to fit the gamma--ray observations of the SNR W28.

In Fig.~\ref{fig:cr_map} we show the CR over--density around the SNR for different values of the particle energy and of the time after the supernova explosion. The upper, middle, and lower panels refer to a time of 6, 19, and 60 kyr after the explosion, respectively. Plots on the first, second, and third column refer to particle energies of $30\,$GeV, $3\,$TeV, and $300\,$TeV, respectively. The escape of CRs is described by Eq.~\ref{eq:escapetime}, which states that higher energy CRs are released first, and lower energy CRs escape at later times. This is the reason why there is no CR excess in the top--left panel of Fig.~\ref{fig:cr_map}: for the choice of parameters made here, for a SNR age of $6\,$kyr particles with an energy of $30\,$GeV are still confined within the SNR shock.
As the age of the SNR increases, CRs diffuse further away along the flux tube and fill a broader and broader region. As a consequence of that, the CR over--density decreases accordingly.
It is evident from these maps that a molecular cloud located in the vicinity of the SNR can be illuminated by the escaping CRs and become a bright gamma--ray source only if it is located within the flux tube. A nearby cloud which is not magnetically connected with the SNR will not be illuminated by CRs, despite its proximity to the SNR.

All the plots in Fig.~\ref{fig:cr_map} refer to a region of size $\approx 200\,$pc around the SNR. As said in Sec.~\ref{sect:model}, this roughly represents the expected length of a magnetic flux tube in the Galaxy \citep[e.g.][]{plesser}. For distances larger than a few hundred parsecs from the SNR, the flux tube loses its identity and it is disrupted due to the exponential divergence of field lines. Thus, the results presented in this paper are accurate and reliable for distances up to a few hundred parsecs and less. 

Since we assumed here that CRs are strongly magnetized, i.e. their remain attached to field lines, their transport across the mean magnetic field is solely governed by the diffusion of field lines, which is described by the diffusion coefficient $D_m$. This quantity determines how quickly the field lines diverge as a function of the displacement $z$ along the mean field. This is demonstrated in Fig.~\ref{fig:Dm}, where the CR over--density for particles of energy $3\,$TeV is shown. The age of the SNR is $19\,$kyr. Three different values for $D_m$ are considered: 0.5, 1, and $2\,$pc for the left, middle, and right panel, respectively. Unfortunately, our knowledge of the properties of the interstellar magnetic field is not good enough to allow a determination or an estimate of this parameter.
For a SNR located in a diffuse interstellar medium (i.e. with no massive molecular clouds) the morphology of the resulting gamma--ray emission due to CR proton--proton interaction with the ambient gas would closely follow the spatial distribution of CRs. 
Thus, observing the diffuse gamma--ray emission generated by runaway CRs around SNRs might serve as a tool to explore the structure of the interstellar magnetic field. The detection of such diffuse emission is within the capabilities of future gamma--ray instruments such as the Cherenkov Telescope Array \citep{acero,sabrinona}.

\begin{figure}
\centering
\includegraphics[width=0.47\textwidth]{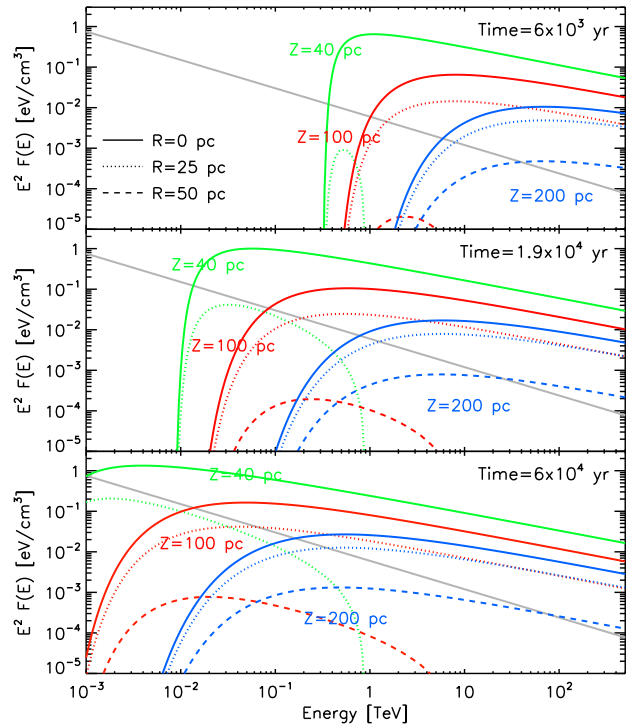}
\caption{Spectra of runaway cosmic rays at different positions and times after the explosion. The age of the supernova remnant is $t=6\,$kyr (upper panel), $t=19\,$kyr (middle panel) and $t=60\,$kyr (lower panel). Solid lines refer to spectra along the $z$-axis, oriented as the mean magnetic field, at three different positions ($z=40\,$pc, $z=100\,$pc and $z=200\,$pc). For each distance $z$, three different values of $R$ (the perpendicular distance from the $z$-axis) are also considered: $R=0\,$pc, $R=25\,$pc, and $R=50\,$pc. The black lines show the CR background.}
\label{fig:spectra}
\end{figure}

Finally, we show in Fig.~\ref{fig:spectra} the spectra of escaping CRs at different distances from the SNR and at different times after the supernova explosion.
Each panel refers to a different epoch: 6, 19, and $60\,$kyr for the top, middle, and bottom panel, respectively. Solid curves show the spectra at three different positions on the $z$-axis: $40\,$pc, $100\,$pc, and $200\,$pc. For each of these positions we also show the spectra at different distances from the $z$-axis: $25\,$pc (dotted lines) and $50\,$pc (dashed lines).

At high energies, in almost all cases the energy spectra are power laws with slope $\approx \alpha+s/2$, where $\alpha$ is the slope of the injection spectrum of runaway CRs and $s$ is the slope of the energy dependent diffusion coefficient (Eq.~\ref{eq:diffusion}). Such a behavior can be inferred from Eq.~\ref{eq:parallel}. If one moves to larger values of $z$, the on--axis (i.e. R = 0) high energy spectrum preserves the same slope, but its normalization decreases as $\approx 1/z$. This is due to the fact that the transverse section of the magnetic flux tube is increasing proportionally to $z$, while the CR intensity along a field line is independent of $z$ for $z^2 \ll 4 D_{\parallel} t$  (see Eqns.~\ref{eq:frw} and \ref{eq:parallel}).
A feature common to all the spectra plotted in Fig.~\ref{fig:spectra} is the presence of a low energy cutoff. The cutoff is due to the fact that at a given time, only particles of sufficiently large energy had enough time to propagate over the distance $z$. The cutoff is moving towards lower energies if a longer time is considered, because particles with lower energies have then the time to reach a given position $z$ along the axis.
Finally, some curves (for example the ones with $z = 40\,$pc and $R = 25\,$pc) exhibit a quite sharp high energy cutoff. This cutoff is due to the fact that particles of different energy are injected within different transverse sections of the flux tube. Higher energy particles are released earlier from the SNR, when the shock radius is smaller, lower energy ones are injected later, when the shock radius is larger. While diffusing along the field lines, CRs are displaced in the transverse direction due to field line wandering. Higher energy particles, which have been injected in a smaller region around $z = 0$, need on average a larger transverse displacement in order to reach a given distance $\hat{R}$ from the $z$--axis. Thus, for small enough $z$, the opening of the magnetic flux tube might not be enough to allow high energy particles to reach $\hat{R}$, and this explains the presence of the cutoff.

\begin{figure}
{\includegraphics[width=.46\textwidth]{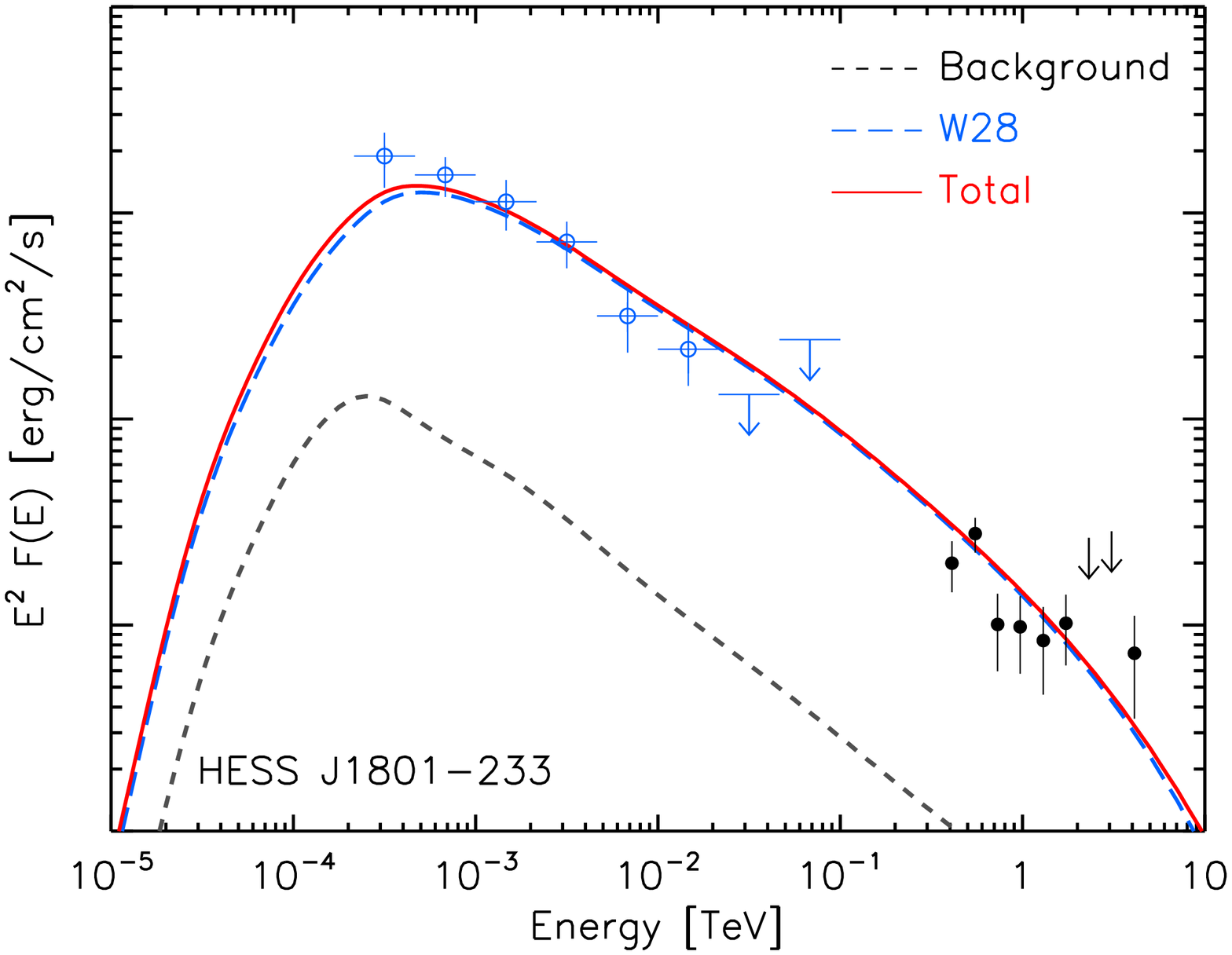}
\includegraphics[width=.46\textwidth]{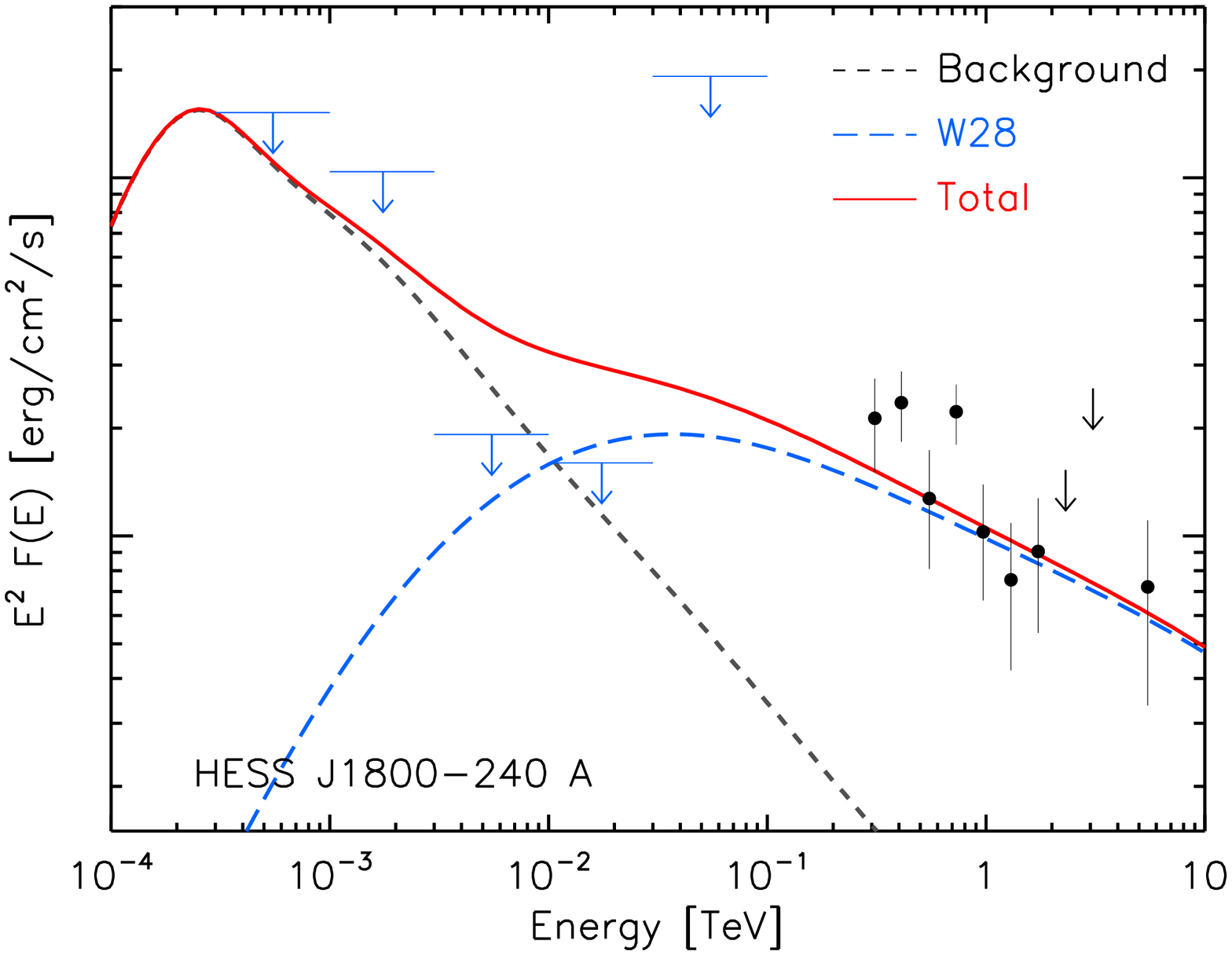}
\includegraphics[width=.46\textwidth]{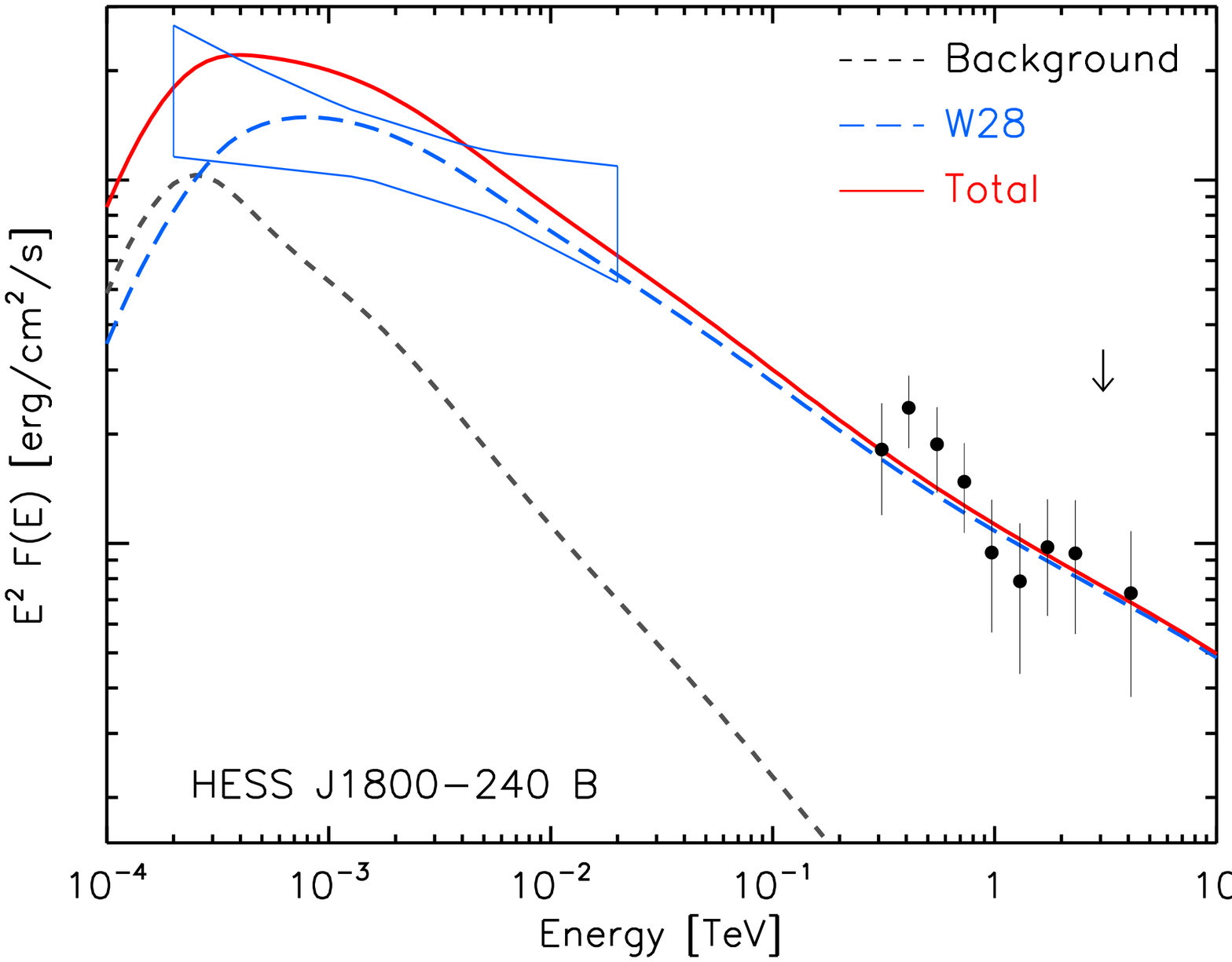}}
\caption{Gamma--ray emission from the three molecular clouds surrounding the supernova remnant W28. Fermi data are shown as open (blue) circles. Filled (black) circles refer to HESS data. The dashed lines show the contribution to the gamma--ray emission from the cosmic ray galactic background, the long-dashed lines show the contribution from cosmic rays that escaped from W28 and the solid (red) line is the total.}
\label{fig:w28}
\end{figure}

In the next Section we apply the model developed above to a specific object, i.e. the SNR W28 and the molecular clouds located in its proximity. Such clouds have been detected in gamma rays and this has been interpreted by many authors as the result of their being illuminated by CRs that escaped the SNR. We will demonstrate that a good agreement can be reached between the predictions of the model and observations and we will discuss the impact of this on the attempts to derive the particle diffusion coefficient close to SNRs by means of gamma--ray observations.

\section{Application to the supernova remnant W28}
\label{sect:W28}

W28 is an old SNR in its radiative phase of evolution, located in a region rich of dense molecular gas with average density $\gtrsim 5\,{\rm cm^{-3}}$. At an estimated distance of $\sim 2\,{\rm kpc}$ the SNR shock radius is $\sim 12\,{\rm pc}$ and its velocity $\sim 80\,{\rm km/s}$ \citep[e.g.][]{rho}. By using the dynamical model by \citet{cioffi} and assuming that the mass of the supernova ejecta is $\sim 1.4\,M_{\odot}$, it is possible to infer the supernova explosion energy ($E_{SN} \sim 0.4 \times 10^{51}{\rm erg}$), initial velocity ($\sim 5500\,{\rm km/s}$), and age ($t_{age} \sim 4.4 \times 10^4 {\rm yr}$).

Gamma ray emission has been detected from the region surrounding W28 both at TeV \citep{W28hess} and GeV energies \citep{W28fermi,W28agile}, by HESS, FERMI, and AGILE, respectively. The TeV emission correlates quite well with the position of three massive molecular clouds, one of which is interacting with the north-eastern part of the shell (and corresponds to the TeV source HESS J1801-233), and the other two being located to the south of the SNR (TeV sources HESS J1800-240 A and B). The masses of these clouds can be estimated from CO measurements and result in $\approx 5$, $6$, and $4 \times 10^4 M_{\odot}$, respectively, and their projected distances from the centre of the SNR are $\approx$ 12, 20, and $20\,$pc, respectively \citep{W28hess}. The GeV emission roughly mimics the TeV one, except for the fact that no significant emission is detected at the position of HESS J1800-240 A.

The gamma--ray emission from the clouds in the W28 region has been interpreted by many authors as the result of the interaction of CRs that escaped W28 with the dense gas in the cloud \citep{fujita,li,gabiciW28,ohira,lazarianW28}. All these approaches started from the assumption of isotropic diffusion of CRs, and a general consensus was found on the fact that, in order to fit observations, the diffusion coefficient had to be suppressed by a factor of $\approx 10 ... 100$ with respect to the average value in the Galaxy, which is $D_{gal} \approx D_0\,(E/10\,{\rm GeV})^{\delta}$ with $D_0 \approx 10^{28}...10^{29}\,$cm$^2$/s and $s \approx 0.3 ... 0.7$ \citep{fiorenzareview}.

In this section we take a different approach and we apply the model developed in Sec.~\ref{sect:model} to estimate the spectrum of CRs and derive the $\gamma$-ray emission expected from the clouds in the W28 region. This approach is radically different from the ones mentioned above because it relies on the more physical assumption that the diffusion of CRs is not isotropic, but proceeds mainly along the magnetic field lines.

The results of our modeling are shown in Fig.~\ref{fig:w28}, where the gamma--ray data from the three molecular clouds are plotted as blue open symbols (data from FERMI) and black filled dots (data from HESS). The emission from the sources  HESS J1801-233, and HESS J1800-240 A and B is plotted in the top, middle, and bottom panel, respectively. The black dashed lines represent the contribution to the gamma--ray emission from the proton--proton interactions of the CRs in the galactic background with the inter--cloud gas. The blue long--dashed lines represents the contribution to the emission from the runaway CRs that escaped from W28. The solid red line is the total emission. The gamma--ray fluxes have been computed following \cite{kamae} with an additional multiplicative factor 1.5 to take into account elements heavier than hydrogen in both cosmic rays and ambient gas \citep{mori}.

A good agreement with observations is obtained is a parallel diffusion coefficient $\widetilde{D}_\parallel = 10^{28}\,$cm$^2$/s with $s = 0.5$ is adopted, together with a diffusion coefficient for field lines $D_m = 1\,$pc with $b^2 = (\delta B/B_0)^2 = 0.2$. Moreover, we assumed that $\approx 20\%$ of the total explosion energy has been converted into CRs with a spectrum proportional to $E^{-2.2}$ and extending from $1\,$GeV to $5\,$PeV. In order to be illuminated by the escaping CRs, the three molecular clouds have to be located in the proximity of the axis of the magnetic flux tube (i.e. the direction of the local mean field). The spectra reported in the figure refers to the positions $z = 10$, $165$, and $35\,$pc and $R = 6.5$, $0$, and $14\,$pc (top to bottom panel, respectively).

It has to be noted that, due to the number of parameters involved in the model, other sets of parameter values might be found that provide an equally satisfactory fit to data. This is not surprising, given that several previous modelings of this source provided an equally good fit to data by using a radically different picture (i.e. isotropic diffusion of CRs) for the transport of particles. Moreover, while a quite small normalization of the (isotropic) diffusion coefficient, roughly of the order of $\widetilde{D} \approx 5 \times 10^{26}$~cm$^2$/s had to be adopted in order to fit data satisfactorily, in the anisotropic case we obtain a good agreement with data for a significantly larger value of the (parallel) diffusion coefficient of $\widetilde{D}_\parallel\approx 10^{28}$~cm$^2$/s. It might be noticed that this number is close to the standard values inferred for the diffusion of CRs in the Galaxy. Thus, any attempt to constrain the CR diffusion coefficient from the observations of gamma--rays from the vicinity of SNRs needs to take into account that an intrinsic uncertainty exists, and it is related to the unknown nature of the CR transport in the interstellar medium, and in particular to the unknown relative relevance of the transport parallel and perpendicular to the magnetic field lines. 

\section{Conclusions}
\label{sect:conclusions}

The details of the transport of CRs in the Galaxy are still little understood. Studies of the composition of CRs provide us with an estimate of the average confinement time of CRs within the Galaxy, which can be translated into a spatially averaged diffusion coefficient for CRs \citep[e.g.][]{andyreview,fiorenzareview}. Whether the CR diffusion coefficient has large spatial variations or it is rather uniform throughout the Galaxy is not known, thought a suppression of diffusion close to CR sources might be expected due to CR streaming instability \citep{plesser,malkov}. To this purpose, the detection of gamma--ray emission from the vicinity of CR accelerators might be used to constrain the CR diffusion coefficient, and thus assess the importance of such suppression \citep[e.g.][]{atoyan,gabici09}. This is because CRs escaping the accelerators would produce gamma rays via proton--proton interactions with the ambient medium. Both the morphology of the resulting emission and its spectrum would depend on the functional form (i.e. energy dependence, level of anisotropy) of the diffusion coefficient.

An object that has been extensively investigated in this context is the SNR W28. Three massive molecular clouds, with total mass in the $\approx 10^5 M_{\odot}$ range, are located in the vicinity of the SNR shell and emit gamma rays. This has been interpreted as the result of the illumination of the clouds by the CRs that escaped the SNR. Several models have been proposed to fit these observations, and all of them are based on the assumption that the diffusion of CRs proceeds isotropically \citep[e.g.][and see Sec.~\ref{sect:W28} for a complete list of references]{W28agile,gabiciW28}. There is a general consensus on the fact that the (isotropic) diffusion coefficient has to be suppressed by a factor of $\approx 10...100$ with respect to the average Galactic one in order to explain the observations. This implies coefficients in the range $\widetilde{D} \approx 10^{26} ... 10^{27}\,$cm$^2$/s.

In this paper, the assumption of isotropy of diffusion has been relaxed, and a more physically motivated situation have been investigated, in which CRs propagate mainly along the magnetic field lines. We considered here the limiting scenario in which the diffusion of CRs across field lines is very small and thus can be neglected. In such a situation, the transverse displacement of CRs is uniquely due to the wandering of the field lines \citep{parker}. Spectra and morphology of the spatial distribution of CRs around SNRs have been computed and described. The main feature is the elongated, filamentary distribution of CRs, as opposed to the spherical distribution found in the case of isotropic diffusion.

In order to fit the gamma--ray data from the W28 region within this scenario, one has to assume that the molecular clouds in its vicinity are magnetically connected to the SNR through a magnetic field flux tube. If this is the case, an accurate fit to data can be obtained. Under this assumption, particles are bound to the flux tube and thus forced to propagate along a specific direction. For plausible values of the diffusion coefficient of magnetic field lines,  in order to obtain the correct CR over--density at the location of the molecular clouds a large (parallel) diffusion coefficient of the order of $\widetilde{D}_\parallel \approx 10^{28}\,$cm$^2$/s has to be adopted.

The fact that a very good agreement has been found with data in the two radically different scenarios characterized by  isotropic and anisotropic diffusion tells us that more data needs to be collected from more SNRs in order to infer with reasonable confidence the properties of the diffusion of particles escaping their accelerators.
The diffuse emission that these runaway particles would produce in their interaction with the ambient gas is, even in the absence of very massive clouds, within the capabilities of the Cherenkov Telescope Array \citep{acero, sabrinona}.
These observations will provide us with precious informations about the properties of the transport of CRs in the Galaxy, but also with a direct evidence for the fact that SNRs are indeed the accelerators of galactic CRs.

\section*{Acknowledgments}

We thank F. Casse, A. Marcowith, F. Piazza, V. Ptuskin, R. Schlickeiser, and L. Sironi for helpful discussions.
The work of LN and SG has been supported by ANR under the JCJC Programme.

\label{lastpage}


\begin{thebibliography}{99}

\bibitem[Abdo et al.(2010)]{W28fermi}
Abdo, A.~A., et al., 2010, ApJ, 718, 348

\bibitem[Abdo et al(2011)]{fermiRXJ}
Abdo, A.~A., et al., 2011, ApJ, 734, 28

\bibitem[Acciari et al.(2011)]{fermitycho}
Acciari, V.,~A., et al., 2011, ApJ Lett., 730, L20

\bibitem[Acero et al.(2012)]{acero}
Acero, F., et al., 2012, accepted to Astropartcle Physics, arXiv:1209.0582

\bibitem[Achterberg \& Ball(1994)]{achterberg}
Achterberg, A., Ball, L., 1994, A\&A, 284, 687

\bibitem[Aharonian \& Atoyan(1996)]{atoyan}
Aharonian, F.~A., Atoyan, A.~M., 1996, A\&A, 309, 917

\bibitem[Aharonian et al.(2008)]{W28hess}
Aharonian, F.~A., et al., 2008, A\&A, 481, 401

\bibitem[Casanova et al.(2010)]{sabrinona}
Casanova, S., Jones, D.~I., Aharonian, F.~A., Fukui, Y., Gabici, S., et al., 2010,  PASJ, 62, 1127

\bibitem[Casse et al.(2002)]{fabien}
Casse, F., Lemoine, M., Pelletier, G., 2002, Phys. Rev. D, 65, 023002

\bibitem[Castellina \& Donato(2012)]{fiorenzareview}
Castellina, A., Donato, F., 2011, arXiv:1110.2981

\bibitem[Chuvilgin \& Ptuskin(1993)]{cptuskin}
Chuvilgin, L.~G., Ptuskin, V.~S., 1993, A\&A, 279, 278

\bibitem[Cioffi et al.(1988)]{cioffi}
Cioffi, D.~F., McKee, C.~F., Bertschinger, E., 1988, ApJ,  334, 252

\bibitem[Drury et al.(1994)]{dav}
Drury, L.~O'C., Aharonian, F.~A., V\"olk, H.~J., 1994, A\&A, 287, 959

\bibitem[Duffy et al.(1995)]{duffy}
Duffy, P., Kirk, J.~G., Gallant, Y.~A., Dendy, R.~O., 1995, A\&A Lett., 302, L21

\bibitem[Ellison et al(2010)]{donRXJ}
Ellison, D.~C., Patnaude, D.~J., Slane, P., Raymond, J., 2010, ApJ, 712, 287 

\bibitem[Fujita et al.(2009)]{fujita}
Fujita, Y., Ohira, Y., Tanaka, S.~J., Takahara, F., 2009, ApJ, 707, L179

\bibitem[Gabici \& Aharonian(2007)]{gabici07}
Gabici, S., Aharonian, F.~A., 2007, ApJ Lett., 665, L131

\bibitem[Gabici et al.(2009)]{gabici09}
Gabici, S., Aharonian, F.~A., Casanova, S., 2009, MNRAS, 396, 1629

\bibitem[Gabici et al.(2010)]{gabiciW28}
Gabici, S., Casanova, S., Aharonan, F.~A., Rowell, G., SF2A--2010: Proceedings of the Annual meeting of the French Society for Astronomy and Astrophysics. Eds.: S. Boissier, M. Heydari-Malayeri, R. Samadi and D. Valls-Gabaud, p.313 -- arXiv:1009.5291

\bibitem[Gabici(2011)]{gabiciescape}
Gabici, S., 2011, MmSAI, 82, 760

\bibitem[Getmantsev(1963)]{getmantsev}
Getmantsev, G.~G., 1963, Soviet Astr., 6, 477

\bibitem[Giacinti et al.(2012)]{giacinti}
Giacinti, G., Kachelriess, M., Semikoz, D.~V., 2012, PRL, 108, 261101

\bibitem[Giordano(2011)]{giordano}
Giordano, F., 2011, MmSAI, 82, 743

\bibitem[Giuliani et al.(2010)]{W28agile}
Giuliani, A., et al., 2010, A\&A, 516, L11

\bibitem[Hillas(2005)]{hillas}
Hillas, A.~M., 2005, J. Phys. G: Nucl. Part. Phys., 31, R95

\bibitem[Hinton \& Hofmann(2009)]{jim}
Hinton, J.~A., Hofmann, W., 2009, ARA\&A, 47, 523

\bibitem[Isichenko(1991)]{isichenko}
Isichenko, M.~B., 1991, Plasma Physics and Controlled Fusion, 33, 795

\bibitem[Jokipii \& Parker(1969)]{parker}
Jokipii, J.~R., Parker, E.~N., 1969, ApJ, 155, 777

\bibitem[Kadomtsev \& Pogutse(1979)]{kadomtsev}
Kadomtsev, B.~B., Pogutse, O.~P., 1979, Nucl. Fusion Suppl., 1, 649

\bibitem[Kamae et al.(2006)]{kamae}
Kamae, T., Karlsson, N., Mizuno, T., Abe, T., Koi, T., 2006, ApJ, 647, 692

\bibitem[Kirk et al.(1996)]{john}
Kirk, J.~G., Duffy, P., Gallant, Y.~A., 1996, A\&A, 314, 1010

\bibitem[Kulsrud \& Pearce(1969)]{kulsrud}
Kulsrud, R., Pearce, W.~P., 1969, ApJ, 156, 445

\bibitem[Li \& Chen(2010)]{li}
Li, H., Chen, Y., 2010, MNRAS, 409, L35

\bibitem[Lingenfelter et al.(1971)]{lingenfelter}
Lingenfelter, R.~E., Ramaty, R., Fisk, L.~A., 1971, Astrophys. Lett., 8, 93

\bibitem[Malkov et al.(2012)]{malkov}
Malkov, M.~A., Diamond, P.~H., Sagdeev, R.~Z., Aharonian, F.~A., Moskalenko, I.~V., arXiv:1207.4728 

\bibitem[Mori(1997)]{mori}
Mori, M., 1997, ApJ, 478, 225

\bibitem[Morlino \& Caprioli(2012)]{giovanni}
Morlino, G., Caprioli, D., 2012, A\&A, 538, A81

\bibitem[Naito \& Takahara(1994)]{nt}
Naito, T., Takahara, F., 1994, J. Phys. G: Nucl. Part. Phys., 20, 477

\bibitem[Ohira et al.(2011)]{ohira}
Ohira, Y., Murase, K., Yamazaki, R., 2011, MNRAS, 410, 1577

\bibitem[Particle Data Group (2008)]{pdg}
Particle Data Group 2008, Phys. Lett. B, 667, 212

\bibitem[Ptuskin et al.(2008)]{plesser}
Ptuskin, V.~S., Zirakashvili, V.~N., Plesser, A.~A., 2008, Adv. Space Res., 42,486 

\bibitem[Rechester \& Rosenbluth(1978)]{rosenbluth}
Rechester, A.~B., Rosenbluth, M.~N., 1978, Phys. Rev. Lett., 40, 38

\bibitem[Rho \& Borkowski(2002)]{rho} 
Rho, J., Borkowski, K.~J., 2002, ApJ, 575, 201

\bibitem[Rodriguez Marrero et al.(2008)]{rodriguez}
Rodriguez Marrero, A.~Y., Torres, D.~F., de Cea del Pozo, E., Reimer, O., Cillis, A.~N., 2008, ApJ, 689, 213

\bibitem[Strong et al.(2007)]{andyreview}
Strong, A.~W., Moskalenko, I.~V., Ptuskin, V.~S., 2007, ARNPS, 57, 285

\bibitem[Torres et al.(2008)]{diego}
Torres, D.~F., Rodriguez Marrero, A.~Y., de Cea del Pozo, E., 2008, MNRAS, 387, L59

\bibitem[Uchiyama et al.(2012)]{W44fermi}
Uchiyama, Y., Funk, S., Katagiri, H., Katsuta, J., Lemoine-Goumard, M., Tajima, H., Tanaka, T., Torres, D.~F., 2012, ApJ Lett., 749, L35 

\bibitem[Yan \& Lazarian(2004)]{lazarian}
Yan, H., Lazarian, A., 2004, ApJ, 614, 757

\bibitem[Yan \& Lazarian(2008)]{lazarian08}
Yan, H., Lazarian, A., 2008, ApJ, 673, 942

\bibitem[Yan et al.(2012)]{lazarianW28}
Yan, H., Lazarian, A., Schlickeiser, R., 2012, ApJ, 745, 140 

\bibitem[Webb et al.(2006)]{webb}
Webb, G.~M., Zank, G.~P., Kaghashvili, E.~Kh., Le Roux, J.~A., 2006, ApJ, 651, 211

\bibitem[Zimbardo et al.(2006)]{zimbardo}
Zimbardo, G., Pommois, P., Veltri, P., 2006, ApJ, 639, L91

\end{thebibliography}
\end{document}